\documentclass{INTERSPEECH2023}

\interspeechcameraready

\usepackage{algorithm}
\usepackage{algpseudocode}
\usepackage{amsmath}
\usepackage{color}
\usepackage[table]{xcolor}
\usepackage{booktabs}
\usepackage{subcaption}
\usepackage{movie15}

\newcommand{\FigTSNE}[2]{
  \begin{subfigure}{0.10\textwidth}
  \includegraphics[width=\textwidth]{images/tsne/#1.pdf}
  \caption{\scriptsize #2}
  \label{fig:#1}
\end{subfigure}}

\newcommand{\FigSpect}[2]{
  \begin{subfigure}{0.21\textwidth}
    \includemovie[poster=images/spect/#1.pdf,mouse,repeat]{\linewidth}{.449691\linewidth}{audios/#1.mp3}
  \caption{#2}
  \label{fig:#1}
\end{subfigure}}

\newcommand{\Href}[1]{
    \href{#1}{#1}
}

\title{SVVAD: Personal Voice Activity Detection for Speaker Verification}
\name{Zuheng Kang, Jianzong Wang$^*$\thanks{$^*$ Corresponding author: Jianzong Wang, jzwang@188.com}, Junqing Peng, Jing Xiao}
\address{Ping An Technology (Shenzhen) Co., Ltd., Shenzhen, China}
\email{\{kangzuheng896, wangjianzong347, pengjq, xiaojing661\}@pingan.com.cn}

\begin{document}

\maketitle

\begin{abstract}
    Voice activity detection (VAD) improves the performance of speaker verification (SV) by preserving speech segments and attenuating the effects of non-speech.
    However, this scheme is not ideal:
    (1) it fails in noisy environments or multi-speaker conversations;
    (2) it is trained based on inaccurate non-SV sensitive labels.
    To address this, we propose a speaker verification-based voice activity detection (SVVAD) framework that can adapt the speech features according to which are most informative for SV.
    To achieve this, we introduce a label-free training method with triplet-like losses that completely avoids the performance degradation of SV due to incorrect labeling.
    Extensive experiments show that SVVAD significantly outperforms the baseline in terms of equal error rate (EER) under conditions where other speakers are mixed at different ratios.
    Moreover, the decision boundaries reveal the importance of the different parts of speech, which are largely consistent with human judgments.
\end{abstract}

\section{Introduction}
\label{sec:introduction}

Voice activity detection (VAD) is a task that identifies whether human speech is present or absent and is often used upstream of other speech components such as automatic speech recognition (ASR), speaker verification (SV), and speaker diarization (SD).
It aims to reduce the impact of non-speech on downstream speech tasks and indirectly improve their performance.
However, their goals are different.
ASR and SD models need to efficiently and accurately determine the boundary between speech and non-speech to avoid missing content.
SV is much more complicated because there are more factors involved.

A typical VAD framework is considered to be a gating module that makes a speech/non-speech decision for each frame.
Early studies focused on signal and statistical analysis and feature engineering \cite{sohn1999statistical,tan2020rvad,kinnunen2013practical,sarkar2022unsupervised}.
More recently, the use of conventional deep learning methods, such as convolutional neural networks (CNN) and recurrent neural networks (RNN), has shown significant improvements in detection performance at low signal-to-noise ratios (SNR) \cite{vafeiadis2019two,wang2019rnn,Dinkel2020}.
Later, with the introduction of the attention mechanism, the model can automatically compare the characteristics of speech and noise within audio to derive more accurate judgments \cite{zheng2020mlnet,kim2018voice}.
Some authors use audiovisual information for VAD detection \cite{ariav2019end,tao2019end}.
Although it improves the performance of SV by VAD to some extent, these frameworks are insufficient because non-target speakers are also identified and retained as speech labels.
Since the SV model uses only a single speaker for supervised training and does not consider multiple speakers, this leads to a significant drop in verification performance.
Therefore, a target speaker-only VAD framework is required.

The personal VAD (PVAD) framework solves this problem by extending the traditional VAD to recognize only the target speaker part and ignore the non-target part.
The feasibility of PVAD has been demonstrated in several studies.
The author of \cite{Ding2020,ding2022personal} proposed and improved the concept of PVAD to solve the problem of ``always running'' models on devices.
\cite{jayasimha2021personalizing,wang2020voicefilter} extends the PVAD to make it easier to use in ASR applications.
Moreover, the concept of PVAD has also been applied to the task of speech enhancement as a secondary task to improve the performance of separation \cite{giri2021personalized,ju2022tea,eskimez2022personalized}.
However, these methods were not applicable to SV.
Although the author of \cite{medennikov2020target,xiao2021microsoft,wang2021bytedance,cheng2023whu,he2021target} tries to employ SD to recognize the speech of different speakers and then find the target speaker by some rules, this type of framework is inefficient for downstream speech tasks due to the complexity of its process.

Traditional PVAD frameworks are inadequate for SV tasks for several reasons:
(1) Traditional PVAD models are trained with frame-by-frame supervision based on human-assigned or ASR forced alignment labels.
However, not every frame predicted by the VAD model has a positive impact on the SV model.
In practice, segments identified as speech by traditional VAD are sometimes associated with low SNR or multiple speakers talking simultaneously, which can severely degrade the performance of SV;
(2) These labels are usually set as hard labels.
However, soft labels are more suitable for SV scenarios because different speech segments contribute differently;
(3) In SV, the VAD model faces a more complex situation and it will be more challenging to take all factors into account.

To address these issues, we have made the following contributions:
(1) We propose a speaker verification-based voice activity detection (SVVAD) framework, which manipulates speech features using FiLM \cite{perez2018film} according to their relevance to the SV model;
(2) We propose a novel label-free training method that uses triplet-like losses to avoid the performance degradation caused by inaccurate human labeling;
(3) Extensive experiments demonstrate that SVVAD achieves significant improvements over the baseline model in terms of equal error rate (EER) under various conditions where other speakers are mixed at different ratios, and that the model-generated VAD decision boundary is highly consistent with human judgment.


\section{Methodology}
\label{sec:method}

\subsection{Recap of Personal VAD Model and Motivation}
\label{sec:recap}

\begin{figure}[t]
    \centering
    \includegraphics[width=0.42\textwidth]{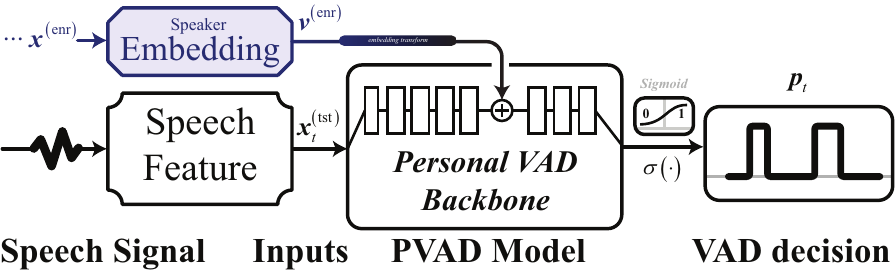}
    \caption{The overview of PVAD framework.}
    \vspace{0.5em}
    \label{fig:pvad}
\end{figure}

To show the advancement of our proposed voice activity detection (VAD) framework, we need to discuss previous approaches.
In the conventional personal VAD (PVAD) framework \cite{Ding2020,ding2022personal,medennikov2020target} (shown in Figure \ref{fig:pvad}), first, a pre-trained speaker verification (SV) model computes the speaker embedding of the enrolled target user $ \boldsymbol{v}^{\left( \mathrm{enr} \right)} $ from his/her recordings.
Then, the PVAD model takes this SV embedding as a priori and the speech features of the audio to be tested $ \boldsymbol{x}_{t}^{\left( \mathrm{tst} \right)} $ as inputs to make predictions $ \boldsymbol{p}_t $ at frame $ t $.
It can be trained using binary cross entropy (BCE) loss.
Eventually, this model is capable of determining whether the frame contains the target speaker.

To further discuss this approach, there is an \textit{Assumption of SV}:
(1) the enrolled speech is restricted to contain only one speaker's voice, and usually stored in the form of speaker embeddings;
(2) the tested audio includes not only the target speaker's speech, but also the non-target speaker's speech and non-speech.
Consequently, the PVAD model is required to identify whether the tested speech contains the target speaker's speech (\texttt{tss}) or its opposite (\texttt{ntss}) with its speaker embedding $ \boldsymbol{v}^{\left( \mathrm{enr} \right)} $, expressed as Equation \ref{eq:old_pvad}.
Where $ \boldsymbol{p}_t= \left[ p_{t}^{\mathtt{tss}},p_{t}^{\mathtt{ntss}} \right] $ can be represented as a posteriori for two categories.
$ \boldsymbol{x}^{\left( \mathrm{tst} \right)}\in \mathbb{R} ^{T\times F} $, $ \boldsymbol{v}^{\left( \mathrm{enr} \right)} \in \mathbb{R} ^{E} $. $ E $, $ F $, $ T $ denote the embedding dimension, feature dimension and sequence length respectively.

\noindent
\begin{equation}
    \label{eq:old_pvad}
    \boldsymbol{p}_t=\mathrm{PVAD}\left( \boldsymbol{x}_{t}^{\left( \mathrm{tst} \right)},\boldsymbol{v}^{\left( \mathrm{enr} \right)} \right)
\end{equation}
\noindent

Although the vanilla PVAD model can identify the target speaker's speech, it is originally designed for non-SV scenarios.
This framework is not optimal for SV because it is based on the hypothesis that retaining only the target speaker's speech and ignoring other sounds can improve the performance of SV.
It is sub-optimal because not all speech of the target speaker contributes positively to SV performance.
Besides, this framework relies heavily on accurate labels.
Therefore, a more advanced VAD model for SV is needed that can account for more complex situations.

\subsection{Improving Personal VAD with Speaker Verification}
\label{sec:svvad}

In order to solve the issue described in $ \S $ \ref{sec:recap}, we propose a speaker verification-based voice activity detection (SVVAD) framework with two novel approaches:
(1) The FiLM layer \cite{perez2018film} is used to construct soft VAD decisions to modify the speech features to automatically identify the most informative parts of the SV model;
(2) A label-free training method is introduced to avoid considering overly complex situations.

\begin{figure}[t]
    \centering
    \includegraphics[width=0.47\textwidth]{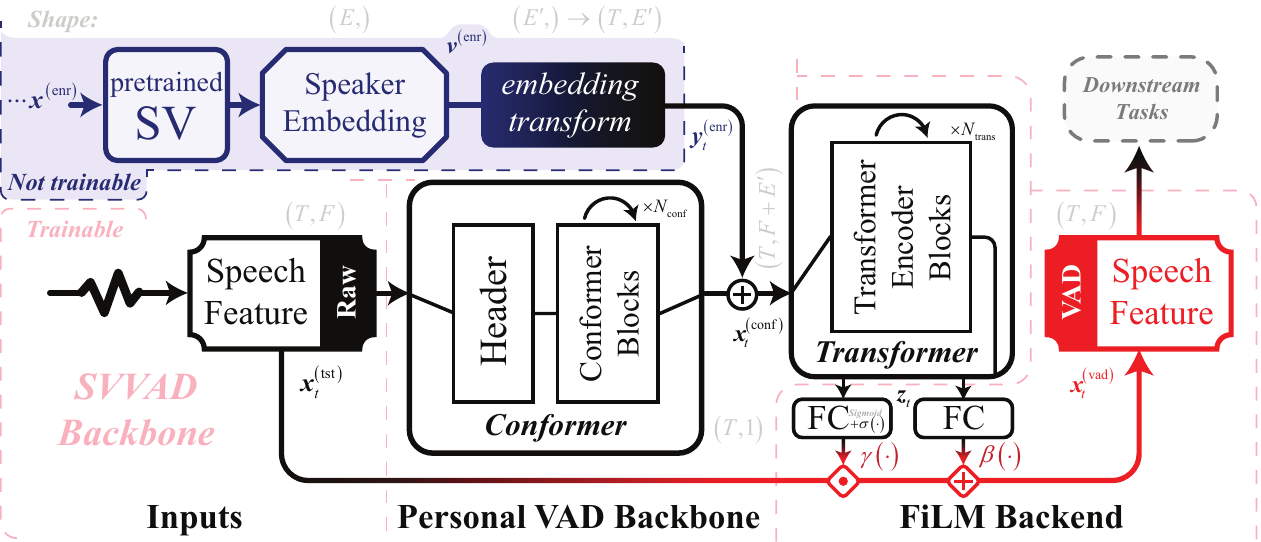}
    \caption{The network architecture of SVVAD backbone.}
    \vspace{-0.5em}
    \label{fig:svvad}
\end{figure}

\noindent
\textbf{Network Architecture for SVVAD Backbone:}
Figure \ref{fig:svvad} outlines the network architecture of the SVVAD backbone.
It consists of a PVAD backbone and a FiLM \cite{perez2018film} backend.
Seen from the input side, this architecture has two branches that receive the speech features to be tested $ \boldsymbol{x}^{\left( \mathrm{tst} \right)} $ and the enrolled target speaker embedding $ \boldsymbol{v}^{\left( \mathrm{enr} \right)} $ inputs respectively.
For the speech feature branch, inspired by \cite{ding2022personal,cheng2023whu,liu2021end}, we use Conformer \cite{gulati2020conformer} (with $ N_{\mathrm{conf}} $ encoder layers) as the speech feature extractor because of its ability to distinguish the features of different speakers more efficiently.
To ensure consistency in the time dimension, we removed the subsampling operation from the original Conformer header.
In this way, both the input $ \boldsymbol{x}_{t}^{\left( \mathrm{tst} \right)} $ and output $ \boldsymbol{x}_{t}^{\left( \mathrm{conf} \right)} $ have the same dimension $ \mathbb{R} ^{T\times F} $.
For the speaker embedding branch, first, a pre-trained SV model is needed to convert the enrolled speech feature $ \boldsymbol{x}^{\left( \mathrm{enr} \right)} $ into speaker embedding $ \boldsymbol{v}^{\left( \mathrm{enr} \right)} $ with dimension $ \mathbb{R} ^{E} $.
Since the embedding size is too large, it can then be reduced to $ \mathbb{R} ^{E'} $ by a fully connected layer.
To merge with another branch, this embedding will be replicated $ T $ times to generate $ \boldsymbol{y}_{t} ^{\left( \mathrm{enr} \right)} $ with dimension $ \mathbb{R} ^{T\times E'} $.

The information from these two branches is then combined by concatenation $ \oplus $, with dimension $ \mathbb{R} ^{T\times \left( E'+F \right)} $.
Since the target speaker's information needs to affect every frame in the tested speech, this problem can be solved by using a self-attentive mechanism, i.e., through several layers of stacked Transformer encoder blocks (with $ N_{\mathrm{trans}} $ layers).
The output $ \boldsymbol{z}_t $ is shown in Equation \ref{eq:transformer}.

\noindent
\begin{equation}
    \label{eq:transformer}
    \boldsymbol{z}_t=\mathrm{Transformer}\left( \boldsymbol{y}_{t}^{\left( \mathrm{enr} \right)}\oplus \boldsymbol{x}_{t}^{\left( \mathrm{conf} \right)} \right)
\end{equation}
\noindent

The fused information $ \boldsymbol{z}_t $ is then sent to two separate fully connected layers (FC) to create VAD decisions with $ \gamma \left( \boldsymbol{z}_t \right) $ and $ \beta \left( \boldsymbol{z}_t \right) $ with dimension $ \mathbb{R} ^{T\times 1} $.
These decisions are then applied to the original speech features $ \boldsymbol{x}_{t}^{\left( \mathrm{tst} \right)} $ by the FiLM layer \cite{perez2018film} to create $ \boldsymbol{x}_{t}^{\left( \mathrm{vad} \right)} $ with dimension $ \mathbb{R} ^{T\times F} $, in Equation \ref{eq:film}.
Since they have different dimensions, it is necessary to use one-to-many operations.

\noindent
\begin{equation}
    \label{eq:film}
    \boldsymbol{x}_{t}^{\left( \mathrm{vad} \right)}=\mathrm{FiLM}\left( \boldsymbol{z}_t \right) =\gamma \left( \boldsymbol{z}_t \right) \cdot \boldsymbol{x}_{t}^{\left( \mathrm{tst} \right)}+\beta \left( \boldsymbol{z}_t \right)
\end{equation}
\noindent

\begin{figure}[b]
    \centering
    \vspace{-0.5em}
    \includegraphics[width=0.47\textwidth]{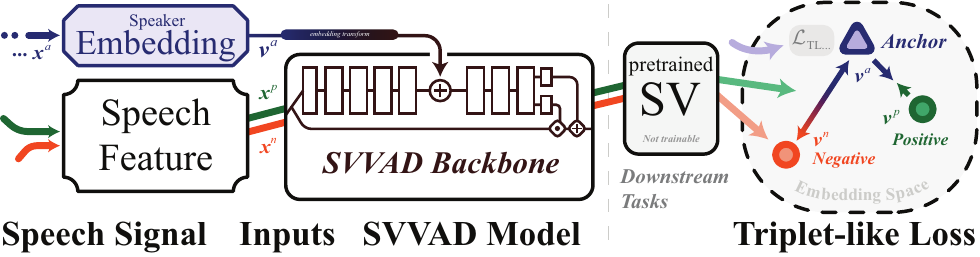}
    \caption{The overview of SVVAD framework.}
    \label{fig:svvad_frame}
\end{figure}

\noindent
\textbf{Label-free Optimization with Triplet-like Loss:}
To completely avoid constructing inaccurate strong VAD labels, it is effective to use only the speaker's label for VAD-label-free training.
Figure \ref{fig:svvad_frame} summarizes the main structure of the SVVAD framework: the modified speech features $ \boldsymbol{x}^{\left( \mathrm{vad} \right)} $ are directly sent into the SV model to obtain the embedding $ \boldsymbol{v}^{\left( \mathrm{vad} \right)} $.
To achieve label-free optimization during training, based on the assumptions of SV, we only need to set different optimization objectives for different cases of the test speech.
Specifically, if the speech to be tested contains the voice of the target speaker, it is considered a positive sample, otherwise it is a negative sample.
This idea is consistent with the triplet-loss (TL) mechanism \cite{schroff2015facenet}, shown in Equation \ref{eq:triplet}.
Where $ \alpha $ is the margin, $ B $ is the batch size.
$ \boldsymbol{x}_{i}^{a} $, $ \boldsymbol{x}_{i}^{p} $ and $ \boldsymbol{x}_{i}^{n} $ is the anchor, positive and negative sample respectively at sample $ i $, and can then be converted to speaker embeddings $ \boldsymbol{v}_{i}^{a} $, $ \boldsymbol{v}_{i}^{p} $ and $ \boldsymbol{v}_{i}^{n} $ by SV model.
$ \lVert \cdot \rVert _{2} $ is 2-norm.
$ \left[ \cdot \right] _+ $ is equivalent to $ \mathrm{max}\left( \cdot,0 \right) $ or $ \mathrm{ReLU}\left(\cdot \right)$.

\noindent
\begin{equation}
    \label{eq:triplet}
    \mathcal{L} _{\mathrm{TL}}=\sum_i^B{\left[ \lVert \boldsymbol{v}_{i}^{a} -\boldsymbol{v}_{i}^{p} \rVert _{2}^{2}-\lVert \boldsymbol{v}_{i}^{a} -\boldsymbol{v}_{i}^{n} \rVert _{2}^{2}+\alpha \right] _+}
\end{equation}
\noindent

Since in SV, scoring uses the cosine distance instead of mean square error (MSE) in Equation \ref{eq:triplet}, the triplet loss can be rewritten in terms of the cosine distance \cite{li2017deep} (denoted as TLcos), in Equation \ref{eq:triplet_cos}.
The cosine distance $ \mathrm{cos}\left(\cdot \right) $ is expressed as 1 minus its cosine similarity.

\noindent
\begin{equation}
    \label{eq:triplet_cos}
    \mathcal{L} _{\mathrm{TLcos}}=\sum_i^B{\left[ \cos \left( \boldsymbol{v}_{i}^{a} ,\boldsymbol{v}_{i}^{p} \right) -\cos \left( \boldsymbol{v}_{i}^{a} ,\boldsymbol{v}_{i}^{n} \right) +\alpha \right] _+}
\end{equation}
\noindent

Since these losses only consider the distance between individual samples and not their overall statistical properties, they may lead to suboptimal performance.
To this end, Lin's concordance correlation coefficient (CCC) \cite{lawrence1989concordance} loss is more reliable and can be used to stabilize the training process, so as to improve the VAD performance.
Then, the Equation \ref{eq:triplet} can be changed to Equation \ref{eq:triplet_ccc} (denoted as TLccc), and Equation \ref{eq:triplet_cos} can be modified to Equation \ref{eq:triplet_cccos}, \ref{eq:triplet_cond} (denoted as TLccos).

\noindent
\begin{equation}
    \label{eq:triplet_ccc}
    \mathcal{L} _{\mathrm{TLccc}}=\left[ \mathrm{CCC} \left( \boldsymbol{v}^{a}, \boldsymbol{v}^{p} \right) -\mathrm{CCC} \left( \boldsymbol{v}^{a} ,\boldsymbol{v}^{n} \right) +\alpha \right] _+
\end{equation}
\noindent

\noindent
\begin{equation}
    \hspace{-2mm}
    \label{eq:triplet_cccos}
    \mathcal{L} _{\mathrm{TLccos}}=\left\{ \begin{matrix}
        \mathrm{CCC}\left( \begin{array}{c}
                                   \cos \left( \boldsymbol{v}^a,\boldsymbol{v}^p \right) +\alpha , \\
                                   \cos \left( \boldsymbol{v}^a,\boldsymbol{v}^n \right)           \\
                               \end{array} \right) & cond>0 \\
        0                                                                  & cond\le 0     \\
    \end{matrix} \right.
\end{equation}
\noindent

\noindent
\begin{equation}
    \label{eq:triplet_cond}
    cond=\cos \left(  \boldsymbol{v}^a, \boldsymbol{v}^p \right) -\cos \left(  \boldsymbol{v}^a, \boldsymbol{v}^n \right) +\alpha
\end{equation}
\noindent

\subsection{Training and Inference}
\label{sec:train_infer}

Let's start by discussing the inference phase, the model first accepts the target speaker's embedding $ \boldsymbol{v}^{\left( \mathrm{enr} \right)} $ and the speech features to be tested $ \boldsymbol{x}^{\left( \mathrm{tst} \right)} $ to construct a VAD decision $ \gamma \left( \boldsymbol{z}_t \right) $ and $ \beta \left( \boldsymbol{z}_t \right) $.
These decisions modify the original speech feature $ \boldsymbol{x}^{\left( \mathrm{tst} \right)} $ by FiLM rules to form $ \boldsymbol{x}^{\left( \mathrm{vad} \right)} $.
It can then be converted into a speaker embedding $ \boldsymbol{v}^{\left( \mathrm{vad} \right)} $ by a pre-trained SV model.
Finally, $ \boldsymbol{v}^{\left( \mathrm{enr} \right)} $ and $ \boldsymbol{v}^{\left( \mathrm{vad} \right)} $ will be scored by cosine similarity.

In the training phase, except for the same operations as in the inference phase, the data-loader will simultaneously construct the anchor, positive and negative samples $ \boldsymbol{x}^{a} $, $ \boldsymbol{x}^{p} $ and $ \boldsymbol{x}^{n} $ in real-time.
These samples are fed into the trainable VAD and non-trainable SV model to obtain the speaker embeddings $ \boldsymbol{v}^{a} $, $ \boldsymbol{v}^{p} $ and $ \boldsymbol{v}^{n} $, which are then trained by the triplet-like losses through Equation \ref{eq:triplet} to \ref{eq:triplet_cccos}.
However, since there is no existing dataset of multi-speaker speech, we will adopt a special method to generate these samples.

\subsection{Training Data Generation}
\label{sec:data_generation}

\begin{algorithm}[t]
    \floatname{algorithm}{Algorithm}
    \renewcommand{\algorithmicrequire}{\textbf{Input:}}
    \renewcommand{\algorithmicensure}{\textbf{Output:}}
    \caption{Training Data Generation Policy}
    \label{peudo:data}
    \scriptsize
    \begin{algorithmic}[1]
        \Require Dataset with single-speaker samples
        \Ensure Multi-speaker samples with anchor, positive, negative
        \For{each iteration in training}
        \State Randomly select the anchor speaker ID as $ d_{a} $
        \State Create anchor samples $ S_{a}\gets \mathrm{get}\left(d_{a}\right) $
        \State Randomly select the number of speakers ($ \le 3 $) for positive or negative, and their corresponding speaker IDs as $ D_{*} $, where $ d_{a}\in D_{p} $ and $ d_{a}\notin D_{n} $
        \State Create empty samples for positive and negative $ S_{*}\gets \phi $
        \While{$ \mathrm{length}\left( S_{*} \right) < $ a fixed duration}
        \State \textcolor[rgb]{0.6,0.6,0.6}{/* the previous and current states are different */}
        \If{$ \mathrm{random}\left(\right) < p_{\mathrm{spk}} $}
        \State $ id\gets \mathrm{select}\left(D_{*}\right)$
        \State $ s_{\mathrm{id}}\gets \mathrm{get}\left(id\right) $
        \If{$ \mathrm{random}\left(\right) < p_{\mathrm{overlap}} $}
        \State $ S_{*}\gets \mathrm{overlapConcat}\left( S_{*},s_{\mathrm{id}} \right) $
        \Else
        \State $ S_{*}\gets \mathrm{concat}\left( S_{*},s_{\mathrm{id}} \right) $
        \EndIf
        \Else
        \State $ s_{\phi}\gets \mathrm{get}\left(\phi\right) $
        \State $ S_{*}\gets \mathrm{concat}\left( S_{*},s_{\phi} \right) $
        \EndIf
        \EndWhile
        \State Audio augmentation for sample $ S_{a} $ and $ S_{*} $
        \EndFor
    \end{algorithmic}
    \Return $ S_{a} $, $ S_{p} $, $ S_{n} $
\end{algorithm}

Multi-speaker samples are generated by concatenating samples from the single-speaker dataset, shown in pseudo-code of Algorithm \ref{peudo:data}.
Where the sign $ * $ is either positive $ p $ or negative $ n $.
$ D_{*} $ is speaker IDs.
$ S_{*} $ is multi-speaker speech samples.
$ p_{\mathrm{spk}} $ and $ p_{\mathrm{overlap}} $ are the probatility of speaking and the speech overlap.
$ \phi $ represents the empty set or the silence.
$ \mathrm{length}\left(\cdot \right) $ is to obtain the length of the sample.
$ \mathrm{random}\left(\right) $ is a value sampled uniformly between 0 and 1.
$ \mathrm{get}\left(\cdot \right) $ is to get audio samples of random duration by the given $ id $ or silence $ \phi $.
$ \mathrm{select}\left(\cdot \right) $ is to randomly select speaker ids.
$ \mathrm{concat}\left(\cdot, \cdot \right) $ is to concatenate the former and the latter to form a longer sequence.
$ \mathrm{overlapConcat}\left(\cdot \right) $ is to operate concatenating but overlapped by a certain ratio.
Audio augmentation is performed as follows: all samples are augmented by probabilistically adding noise with SNR of 10 dB to 30 dB, and RIR reverberation.
With this approach, the anchor, positive and negative samples ($ S_{a} $, $ S_{p} $ and $ S_{n} $) are created.
Since the SV model is well trained and SVVAD requires an explicit learning target, the sample $ \boldsymbol{x}^{a} $ is used directly to construct $ \boldsymbol{x}^{p} $.

\section{Experiments}
\label{sec:experiments}

\subsection{Experimental Setup}
\label{sec:setup}

To validate our proposed SVVAD framework, we employ a pre-trained SV model\footnote[1]{\Href{https://huggingface.co/speechbrain/spkrec-ecapa-voxceleb}} that shares the same experimental setup with our framework.
That is, the SV model uses ECAPA-TDNN \cite{DesplanquesTD20} from SpeechBrain \cite{speechbrain} trained on the Voxceleb dataset \cite{Nagrani19}.
The SVVAD has trained with Voxceleb 1+2 dataset and tested on the test set constructed from the Voxceleb1 (cleaned) verification set by concatenating audio segments from different speakers with varying proportions $ \mathcal{P} $ (percentage of the duration of other speakers' voices).
Considering the variability inherent in the generation of synthetic data, each set of experimental data will be constructed 3 times.
The test metrics are similar to the SV frameworks and are reported as the equal error rate (EER) in percentage (\%) and minimum decision cost function (minDCF) at $ P_{\mathrm{target}}=0.01 $ with $ C_{\mathrm{FA}}=C_{\mathrm{Miss}}=1 $ \cite{DesplanquesTD20}.

Two baseline methods were added to the experiment for comparison to assess the impact of the different methods on SV performance:
(1) WebRTC-VAD \cite{webrtcvad} with default settings;
(2) The traditional PVAD model (in $ \S $ \ref{sec:recap}).
For the SVVAD framework, the SV performance of the four variants of the triplet-like loss (in Equation \ref{eq:triplet} to \ref{eq:triplet_cond}) will be measured.

\vspace{-0.4em}
\subsection{Implementation Details}
\label{sec:implement}

For speech features, the entire framework uses the same Log-Mel-Filterbanks extractor as the pre-trained SV model \cite{speechbrain}.
In SVVAD, the size of the Conformer encoder is $ F=256 $.
The number of Conformer and Transformer layers in SVVAD is $ N_{\mathrm{conf}}=4 $ and $ N_{\mathrm{trans}}=3 $ respectively, and both have a feed-forward size of 256.
The size of the speaker embedding and the shrunken embedding are $ E=192 $ and $ E'=64 $ respectively.
The parameter $ \alpha $ in triplet-like loss is 0.9, 0.5, 0.55, 0.55 in TL, TLcos, TLccc and TLccos respectively.
The probability $ p_{\mathrm{spk}}=0.9 $ and $ p_{\mathrm{overlap}}=0.3 $.

In the first stage, the model is optimized by SGD optimizer for fast convergence, with the learning rate of 1e-2, the momentum of 0.9, and the weight decay of 4e-4.
The speech durations of training of anchor, positive and negative are 6, 8, 8 seconds.
The batch size is $ B=8 $.
In the fine-tuning stage, the optimizer is switched to AdamW \cite{loshchilov2017decoupled}, and the learning rate is changed to 1e-4, and the weight decay is 2e-5.
The speech durations of these three are changed to 8, 12, 12 seconds.
The batch size is $ B=64 $ with gradient accumulation.

\subsection{Evaluation Results}
\label{sec:results}

\begin{table}[t]
    \centering
    \scriptsize
    \setlength{\tabcolsep}{4.1pt}
    \renewcommand{\arraystretch}{1.0}
    \caption{Overall comparison of the proposed SVVAD with 4 triplet-like losses and the baseline models on EER (\%) and minDCF (denoted as $ C_{0.01} $). + denotes adding methods.}
    \begin{tabular}{@{}lcccccccc@{}}
        \toprule
        \multicolumn{1}{c}{$ \mathcal{P} $} & \multicolumn{2}{c}{0\%} & \multicolumn{2}{c}{30\%} & \multicolumn{2}{c}{50\%} & \multicolumn{2}{c}{70\%}                                                                      \\ \cmidrule(l){2-9}
                                            & EER                     & $ C_{0.01} $             & EER                      & $ C_{0.01} $             & EER           & $ C_{0.01} $    & EER            & $ C_{0.01} $    \\ \midrule
        \multicolumn{9}{l}{\textbf{\textit{Baselines}}}                                                                                                                                                                     \\
        w/o VAD                             & \textbf{0.90}           & \textbf{0.1104}          & 2.27                     & 0.2284                   & 14.84         & 0.6950          & 34.70          & 0.9942          \\
        WebRTC                              & 1.14                    & 0.1215                   & 2.31                     & 0.2332                   & 14.50         & 0.6733          & 33.94          & 0.9873          \\
        PVAD                                & 6.22                    & 0.5084                   & 7.95                     & 0.5621                   & 9.89          & 0.6080          & 16.29          & 0.7893          \\ \midrule
        \multicolumn{9}{l}{\textbf{\textit{SVVAD + Triplet-like Loss (ours)}}}                                                                                                                                              \\
        +TL                                 & 4.63                    & 0.4412                   & 6.39                     & 0.4794                   & 8.62          & 0.5893          & 18.32          & 0.8621          \\
        +TLcos                              & 5.48                    & 0.5253                   & 7.15                     & 0.5336                   & 9.73          & 0.6012          & 21.37          & 0.9157          \\
        +TLccc                              & 1.41                    & 0.1201                   & \textbf{1.89}            & \textbf{0.1996}          & \textbf{6.91} & \textbf{0.4912} & \textbf{13.42} & \textbf{0.7396} \\
        +TLccos                             & 2.06                    & 0.1926                   & 4.30                     & 0.3729                   & 7.26          & 0.5433          & 16.30          & 0.7605          \\ \bottomrule
    \end{tabular}
    \vspace{-1em}
    \label{tab:all}
\end{table}

\begin{table}[t]
    \centering
    \scriptsize
    \setlength{\tabcolsep}{14.7pt}
    \renewcommand{\arraystretch}{1.0}
    \caption{Ablation study of the effect of the combination of two types of losses on SV performance on EER (\%).}
    \begin{tabular}{lcccc}
        \toprule
        $ \mathcal{L} $ / $ \mathcal{P} $ & 0\%           & 30\%          & 50\%          & 70\%           \\ \midrule
        TL                                & 4.63          & 6.39          & 8.62          & 18.32          \\
        +TLcos                            & 4.17          & 5.76          & 7.75          & 17.29          \\ \midrule
        TLccc                             & 1.41          & 1.89          & 6.91          & 13.42          \\
        \rowcolor{black!10}
        +TLccos                           & \textbf{1.18} & \textbf{1.70} & \textbf{5.81} & \textbf{12.20} \\ \bottomrule
    \end{tabular}
    \label{tab:ablation}
    \vspace{-1.5em}
\end{table}

\textbf{Overall Evaluation:}
Table \ref{tab:all} compares the EER and minDCF of various methods at different $ \mathcal{P} $.
As for the baseline, the WebRTC-VAD does not enhance the SV performance, sometimes even worse.
The PVAD model can improve the SV performance to some extent, but the improvement is limited.
For SVVAD, among 4 triplet-like losses, the addition of CCC leads to a significant improvement in the SV performance.
But surprisingly, the cosine distance-based loss, which is theoretically more suitable because it is used for scoring, perform worse than the other losses.
In our analysis, this may be due to the fact that the cosine distance-based loss considers only one value as the learning target instead of using the $ E $-dimensional embedding for optimization, which may lead to a large amount of loss of embedding information when training SVVAD, thus making it difficult to converge.
Furthermore, there is a noticeable drop in SV performance when no one else is speaking ($ \mathcal{P}=0\% $), which is unsatisfactory.

\noindent
\textbf{Ablation Study:}
To solve the above problem, we combine two types of losses, which can not only effectively reduce the ambiguity of the learning objectives, but also optimize the cosine distance for scoring.
The experimental results are reported in Table \ref{tab:ablation}.
For faster training, the model is trained on the basis of TL and TLccc.
The total loss $ \mathcal{L} $ consists of two sub-losses, they are $ \left(\mathcal{L}_{\mathrm{TLcos}}+\xi \cdot \mathcal{L}_{\mathrm{TL}}\right) $ or $ \left(\mathcal{L}_{\mathrm{TLccos}}+\xi \cdot \mathcal{L}_{\mathrm{TLccc}}\right) $, where $ \xi $ is the hyperparameter determined by the experiment (set as 0.1).
The results demonstrate that by integrating these two sub-losses, the SV performance is further improved when $ \mathcal{P} $ is larger.
In addition, there is no major performance degradation at $ \mathcal{P}=0\% $.
The best-performing model achieves relative EER reductions over PVAD of 78.6\%, 41.3\% and 25.1\% when $ \mathcal{P} $ is 30\%, 50\% and 70\% respectively.

\begin{figure}[t]
    \centering
    \caption{The speech features of $ \boldsymbol{x}^{\left( \mathrm{tst} \right)} $ and $ \boldsymbol{x}^{\left( \mathrm{vad} \right)} $ with their VAD decision boundaries
    \textcolor[rgb]{0.6,0.6,0.6}{(click on the figure to hear the sound)}.}
    \FigSpect{xcat}{Non-target Speaker}
    \FigSpect{xove}{Noise}
    \label{fig:spect}
\end{figure}

\begin{figure}[t]
    \centering
    \caption{The t-SNE plots of $ \boldsymbol{v}^{\left( \mathrm{vad} \right)} $ for 10 example speakers in conditions of (1) with or without SVVAD; (2) the different $ \mathcal{P} $.}
    \FigTSNE{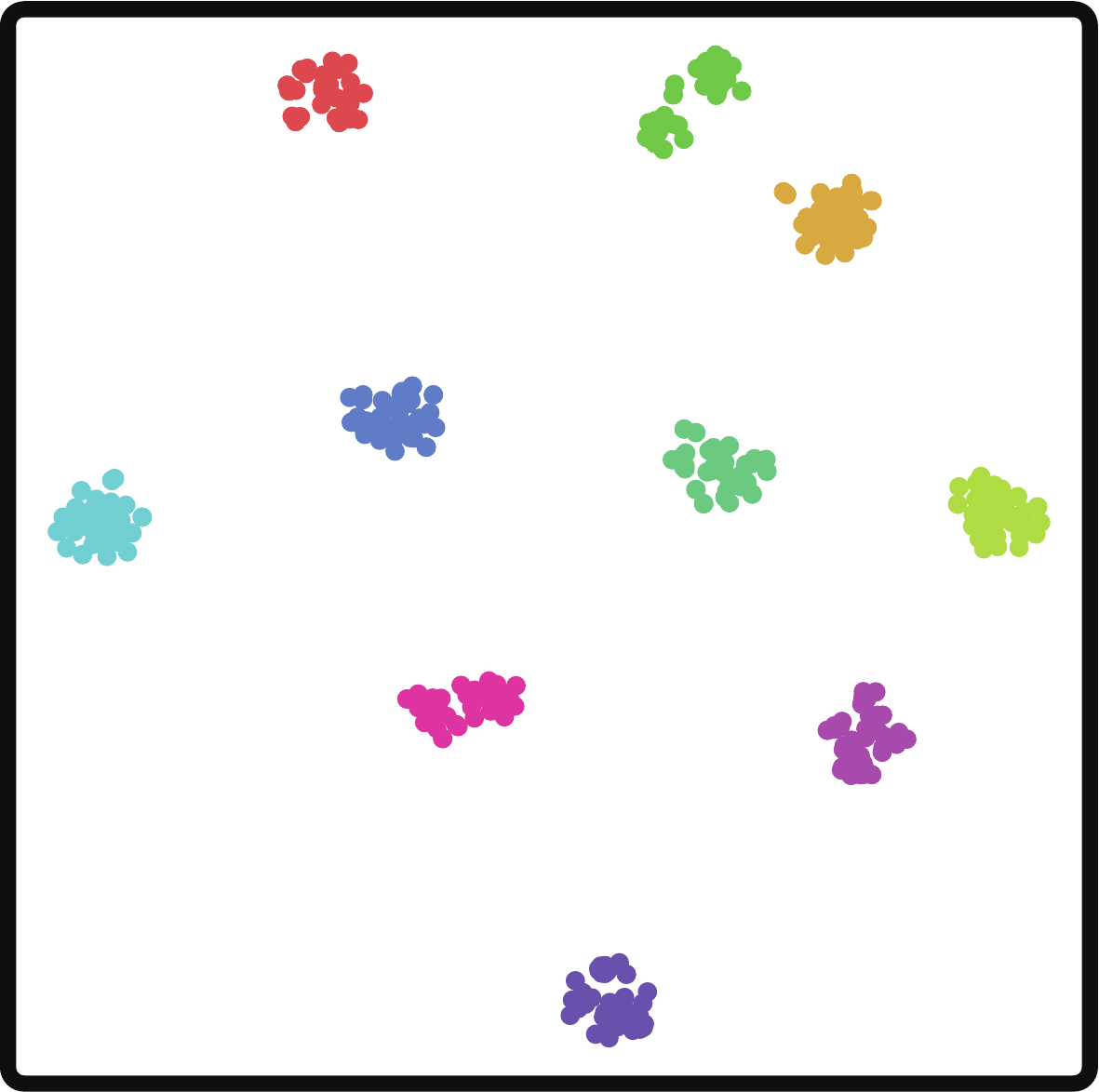}{w/o; 0\%}
    \FigTSNE{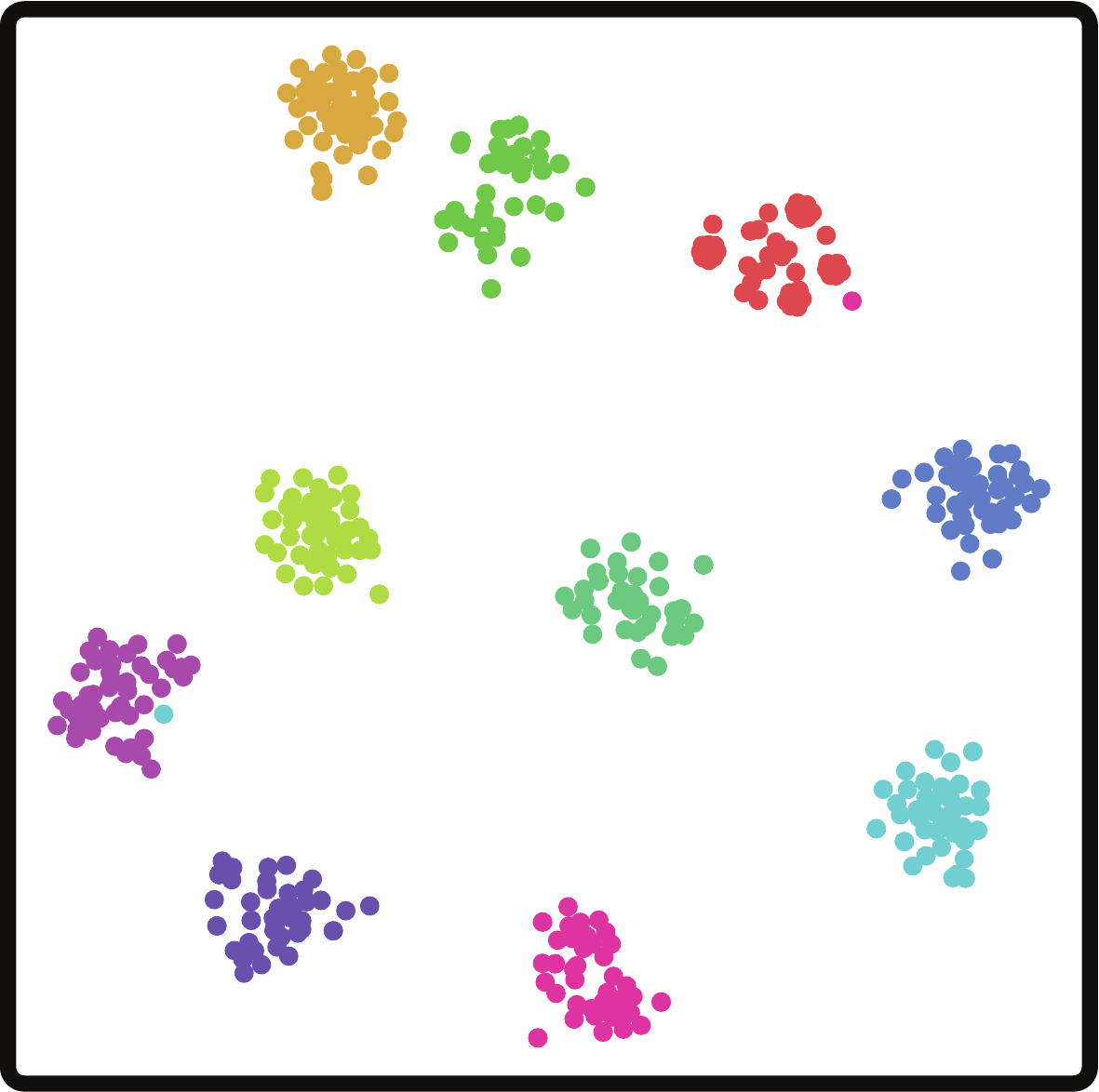}{w/o; 30\%}
    \FigTSNE{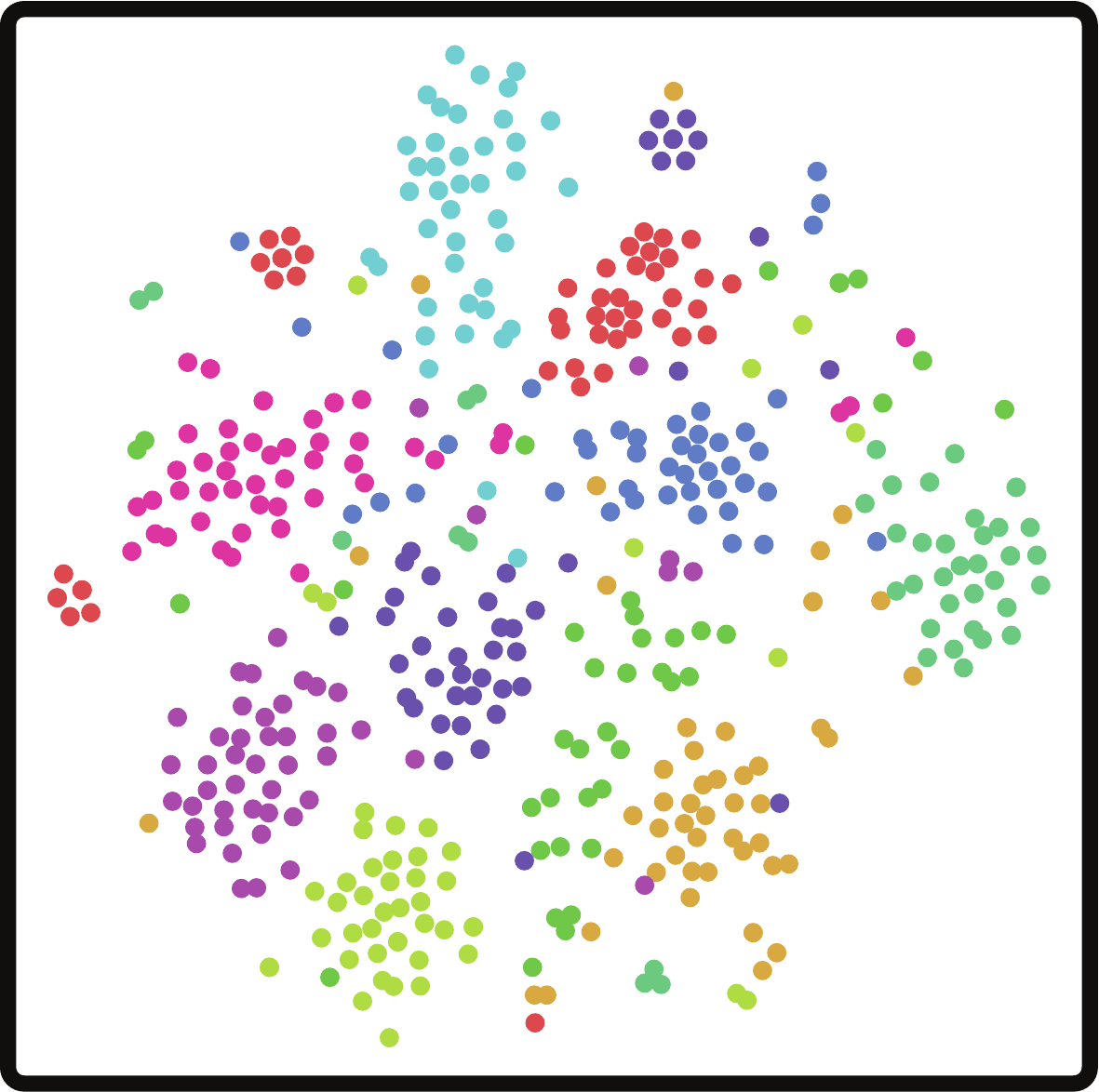}{w/o; 50\%}
    \FigTSNE{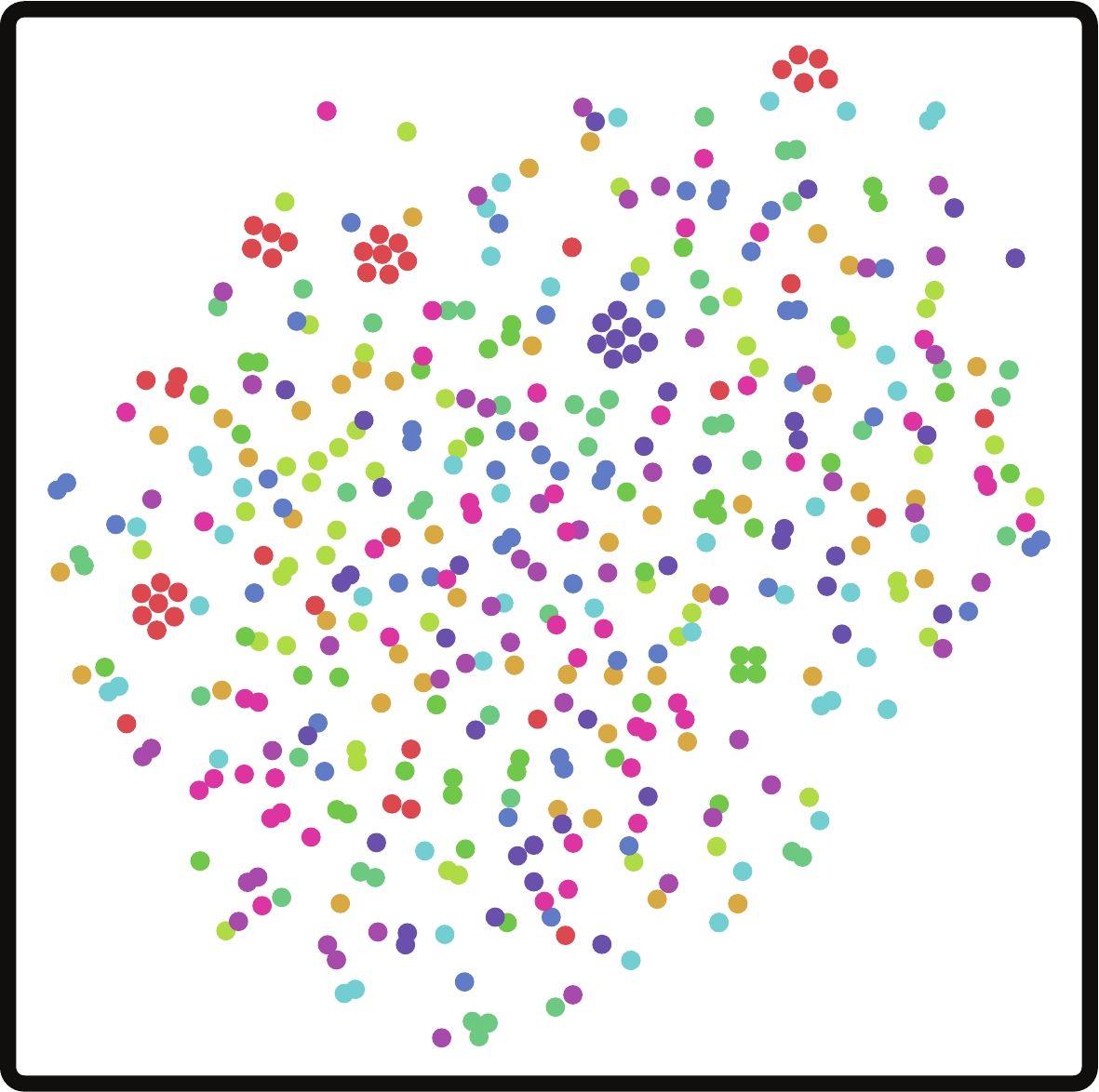}{w/o; 70\%}
    \FigTSNE{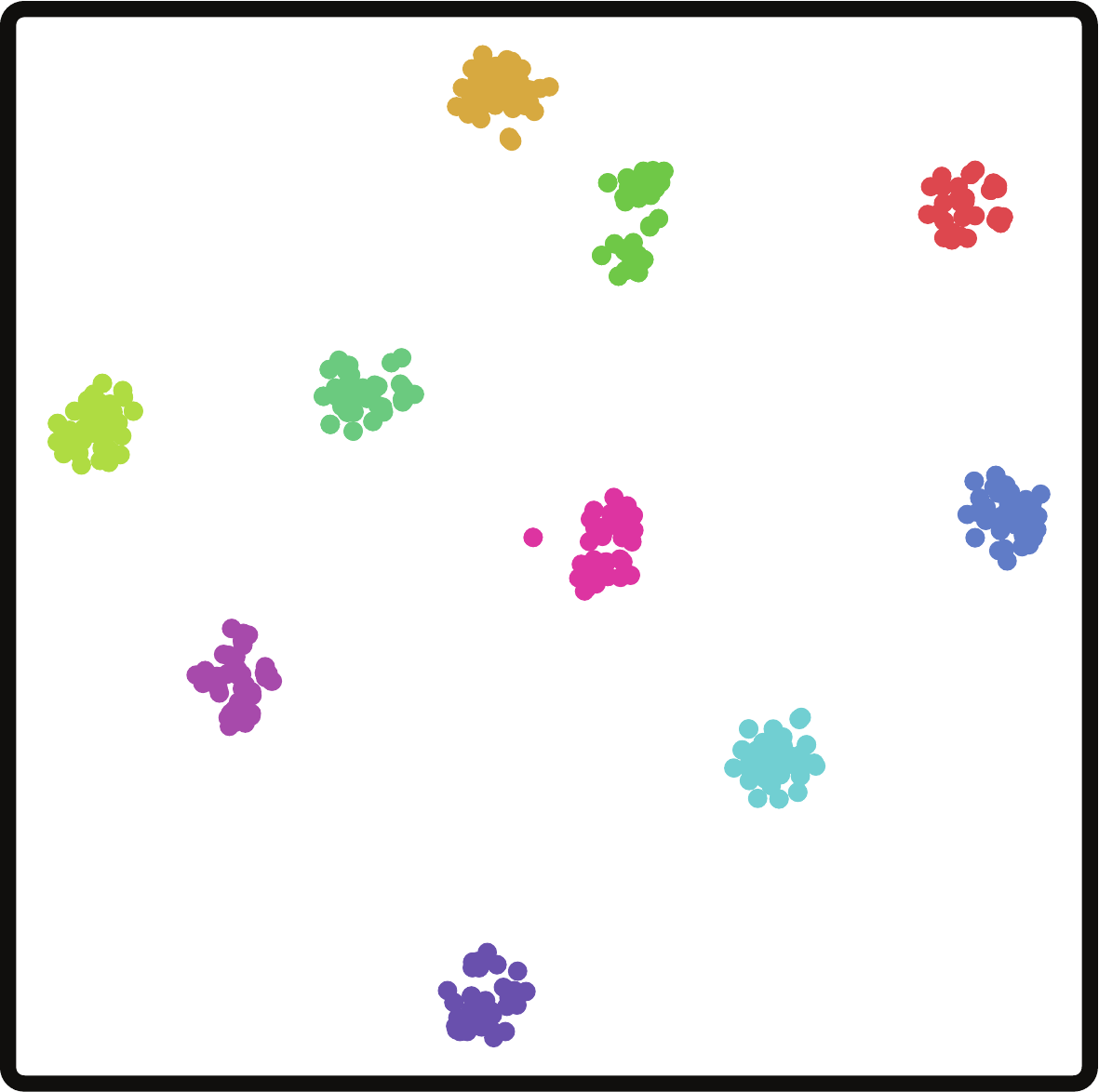}{w/; 0\%}
    \FigTSNE{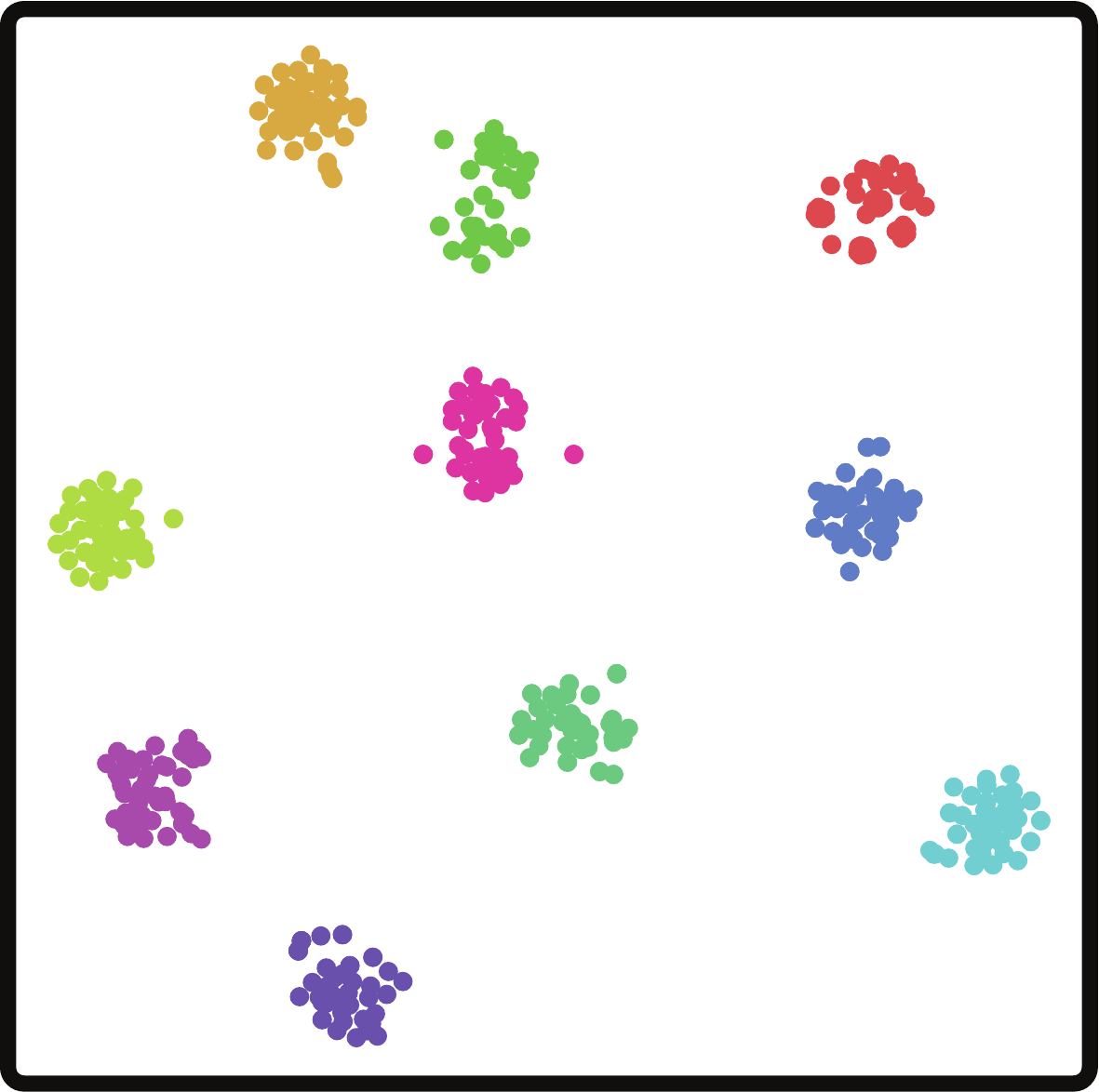}{w/; 30\%}
    \FigTSNE{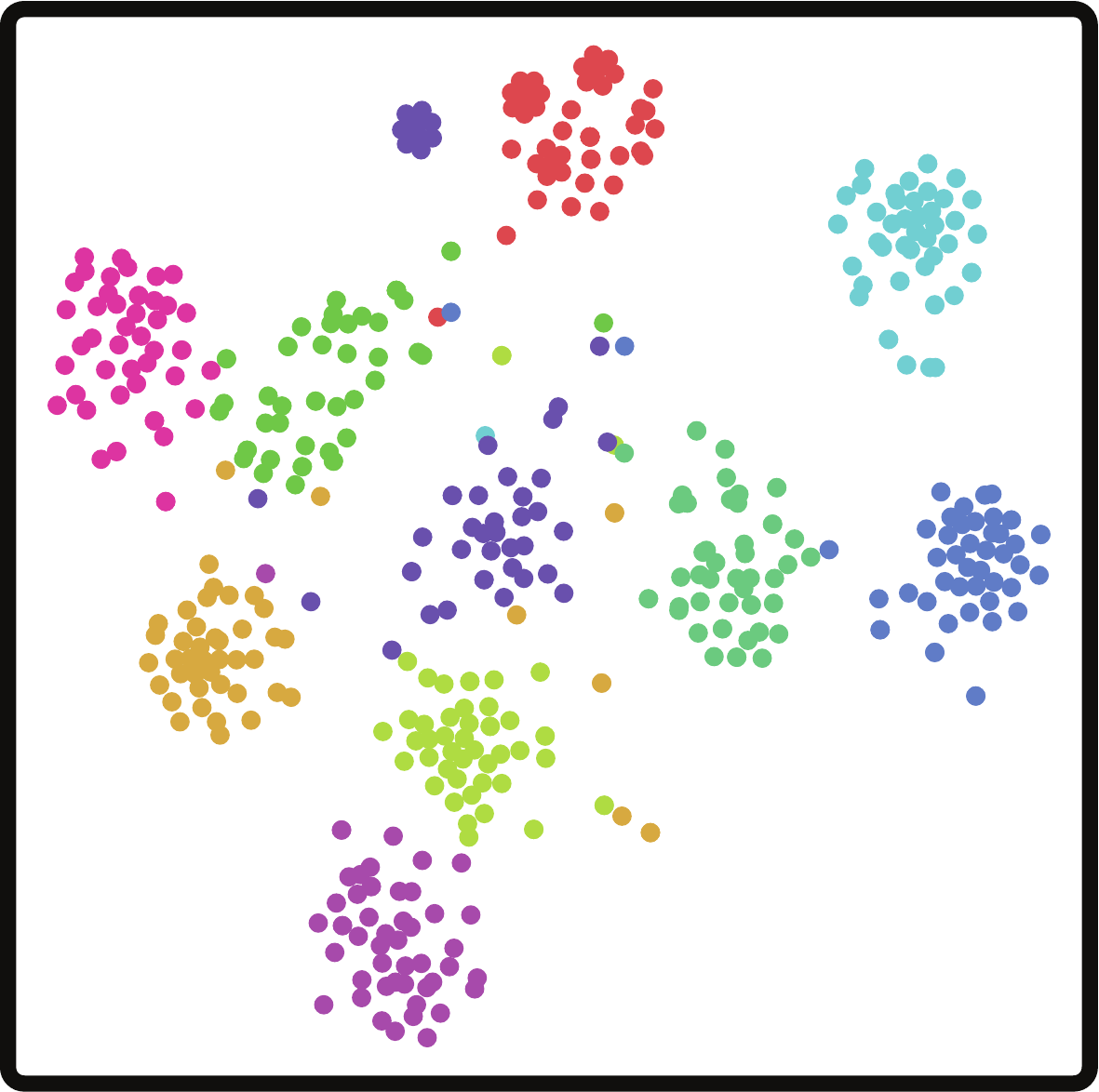}{w/; 50\%}
    \FigTSNE{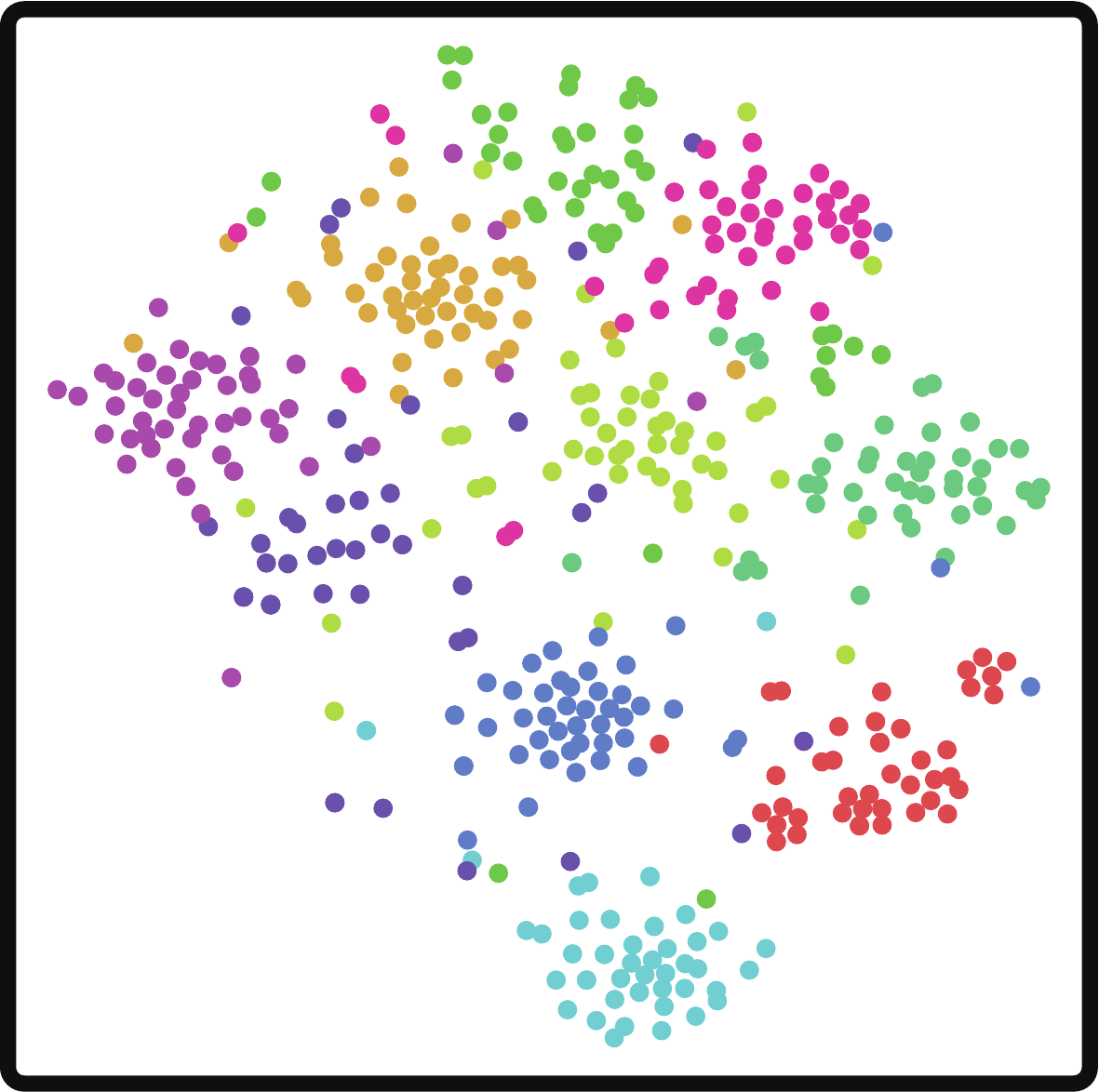}{w/; 70\%}
    \vspace{-1em}
    \label{fig:tsne}
\end{figure}

\noindent
\textbf{Generated Samples:}
According to Figure \ref{fig:tsne}, SVVAD produces more aggregated speaker embeddings when $ \mathcal{P} $ is large, indicating that the model can achieve a greater improvement in SV performance when the input is mixed with more speech from non-target speakers.
Conversely, when $ \mathcal{P} $ is small, the compactness of the cluster looks similar, suggesting that the SV performance is nearly the same.
Figure \ref{fig:spect} illustrates the decision boundaries generated by SVVAD, indicating that, similar to human perception, the model attempts to mask out the speech of non-target speakers and reduce strong noise, thereby suppressing their impact on SV performance degradation.
However, the embient noise may be negligible probably because the SV model is inherently robust to it.

\section{Conclusions}
\label{sec:conclusions}

In this paper, we present SVVAD, a speaker verification-based voice activity detection framework according to which speech features are most beneficial for SV.
We also introduce a label-free training method that optimizes with the triplet-like loss approach, without relying on human labeling.
Extensive experiments on various scenarios show that SVVAD outperforms the baselines in terms of EER in noisy and multi-speaker conditions.
Furthermore, the decision boundaries reveal the importance of different parts of speech for SV, which is largely consistent with human perception.
This work opens up new possibilities for developing more robust and accurate VADs for SV systems in real-world applications.

\section{Acknowledgements}
\label{sec:acknowledgements}

Supported by the Key Research and Development Program of Guangdong Province (grant No. 2021B0101400003) and Corresponding author is Jianzong Wang (jzwang@188.com).


\bibliographystyle{IEEEtran}

\bibliography{ref}

\end{document}